\documentstyle[12pt]{article}

\newcommand{\be}{\begin{equation}}
\newcommand{\ee}{\end{equation}}
\newcommand{\bea}{\begin{eqnarray}}
\newcommand{\eea}{\end{eqnarray}}

\newcommand{\Int}{\int\limits}
\newcommand{\ep}{\varepsilon}
\newcommand{\Li}[2]{{\mbox{Li}}_{#1}\left(#2\right)}

\def\thebibliography#1{\centerline{\bf References}
\list
 {[\arabic{enumi}]}{\settowidth\labelwidth{[#1]}\leftmargin\labelwidth
 \advance\leftmargin\labelsep
 \usecounter{enumi}}
 \def\newblock{\hskip .11em plus .33em minus .07em}
 \sloppy\clubpenalty4000\widowpenalty4000
 \sfcode`\.=1000\relax}

\newcommand{\bi}[1]{\vspace{-3mm} \bibitem{#1}}

\begin{document}
 \thispagestyle{empty}
 \begin{flushright}
 {\tt University of Bergen, Department of Physics}    \\[2mm]
 {\tt Scientific/Technical Report No.1994-17}    \\[2mm]
 {\tt ISSN 0803-2696} \\[9mm]
 {December 1994}           \\
\end{flushright}
 \vspace*{3cm}
 \begin{center}
 {\bf \large
Two-loop three-point diagrams with irreducible numerators
}
 \end{center}
 \vspace{1cm}
 \begin{center}
 Natalia~I.~Ussyukina$^{a,}$\footnote{E-mail address:
                                      ussyuk@theory.npi.msu.su}
 \ \ and \ \
Andrei~I.~Davydychev$^{b,}$\footnote{On leave from
 Institute for Nuclear Physics,
 Moscow State University, 119899, Moscow, Russia.\\
 $\hspace*{6mm}$ E-mail address: davyd@vsfys1.fi.uib.no}
\\
 \vspace{1cm}
$^{a}${\em
 Institute for Nuclear Physics, Moscow State University, \\
 119899, Moscow, Russia
     }
\\
\vspace{.3cm}
$^{b}${\em Department of Physics, University of Bergen,\\
      All\'{e}gaten 55, N-5007 Bergen, Norway}
 \end{center}
 \hspace{3in}
 \begin{abstract}
We study the problem of calculating two-loop three-point diagrams with
irreducible numerators (i.e. numerators which cannot
be expressed in terms of the denominators).
For the case of massless internal particles and arbitrary
(off-shell) external momenta, exact results are obtained in terms
of polylogarithms.
We also consider the tensor decomposition of two-loop three-point diagrams,
and show how it is connected with irreducible numerators.
 \end{abstract}

\newpage
 \setcounter{page}{2}

{\bf 1.} The study of two-loop diagrams is required by increasing precision
of experiments testing QCD and the Standard Model. As compared with
the one-loop case, the calculation of two-loop diagrams is technically
much more complicated. Exact results are known
for a few special cases only, mainly for two-point functions.
There are, however, some approaches which make it possible to obtain
numerical results \cite{Num1,FT} or analytic expansions \cite{BDST}
providing reasonable precision for the cases of interest.

Three-point two-loop diagrams are much less investigated. In this case,
we have two independent external momenta (from which three independent
invariants can be constructed). Although some of the numerical approaches
mentioned above (see also in \cite{Num2}) can also be applied to three-point
functions, this usually requires a larger number of
parametric integrations to be done numerically, and the structure of
the singularities in the integrand is more complicated.
Asymptotic expansions can also be constructed, but they involve
three-fold series in external momentum invariants (see, e.g., in \cite{FT}).
Furthermore, the expansion at large external momenta
requires more information about three-point diagrams with massless
internal lines involving not only higher powers of the propagators,
but also some numerators which cannot be cancelled against any of
the denominators. We shall refer to these as ``irreducible numerators''.
One also obtains them in realistic calculations
with vector and spinor particles. Moreover, the
problem of tensor decomposition of integrals with uncontracted
Lorentz indices is also closely connected with irreducible numerators.
This is one of the essential
complications of two-loop three-point calculations as compared with
both one-loop case \cite{PV,PLB'91} and two-point self-energies \cite{WSB}.

The problem of irreducible numerators (and the related problem of tensor
reduction) is the main subject to be discussed in the present paper.
In fact, it can be considered as a continuation of the
papers \cite{UD1,UD3} where some exact results for three-point
two-loop diagrams were derived\footnote{In the papers \cite{UD2,Bro-lad}
some results
were generalized to the case of an arbitrary number of loops.}.
The remainder of the paper is organized as follows. In section~2
we collect some useful results for one-loop triangle diagrams.
In section~3 we present some results for two-loop three-point
functions and discuss the problem of irreducible numerators. Then,
we calculate integrals with irreducible numerators for the planar
(section~4) and non-planar (section~5) cases.
In section~6 we consider the tensor decomposition
of three-point two-loop diagrams. In section~7 we discuss the results.

\vspace{5mm}

{\bf 2.} Here we shall briefly summarize
some useful formulae for one-loop integrals.

{\em Definition.} The $n$-dimensional one-loop three-point
Feynman integral is defined as
\be
\label{defJ}
J (n; \; \nu_1  ,\nu_2  ,\nu_3 | \; p_1^2, p_2^2, p_3^2 ) \equiv \int
 \frac{\mbox{d}^n q}{ ((p_2 -q )^2)^{\nu_1}  ((p_1 +q )^2)^{\nu_2}
      (q^2)^{\nu_3} } ,
\ee
where all external momenta $p_1, p_2, p_3$ (such that $p_1+ p_2+ p_3=0$)
are ingoing, and the ``causal'' prescription
$(q^2)^{-\nu} \leftrightarrow (q^2+\mbox{i}0)^{-\nu}$ is understood.
Below we shall omit the arguments $p_1^2, p_2^2, p_3^2$ in $J$.

{\em ``Uniqueness'' relations.} At special values of the
sum of the powers of denominators, $\sum \nu_i \equiv \nu_1+\nu_2+\nu_3$,
the following formulae \cite{VPH,UsKaz'83} are valid:
\be
\label{uniq1}
\left. \frac{}{}
J (n; \; \nu_1,\nu_2,\nu_3) \right|_{\Sigma \nu_i = n}
=  \pi^{n/2} \; \mbox{i}^{1-n} \prod_{i=1}^{3}
\frac{\Gamma ( n/2 - \nu_i)}{\Gamma (\nu_i)} \;
\frac{1}{(p_i^2)^{n/2-\nu_i}} ,
\ee
\bea
\label{uniq3}
\left\{ \left.   \frac{}{}
\nu_1 J (n; \; \nu_1 +1,\nu_2,\nu_3)
+ \nu_2 J (n; \; \nu_1 ,\nu_2 +1,\nu_3)
+ \nu_3 J (n; \; \nu_1 ,\nu_2,\nu_3 +1) \right\}
\right|_{\Sigma \nu_i = n-2}
\nonumber \\
=  \pi^{n/2} \; \mbox{i}^{1-n} \prod_{i=1}^{3}
\frac{\Gamma ( n/2 - \nu_i -1)}{\Gamma (\nu_i)} \;
\frac{1}{(p_i^2)^{n/2-\nu_i-1}}.
\hspace{1cm}
\vspace{-3mm}
\eea
Note that the sum of the powers of inverse momenta squared
on the r.h.s. of (\ref{uniq1}) is $n/2$.

{\em The case} $\nu_1=\nu_2=\nu_3=1, \; n=4$. The result for this case is
well-known (see, e.g., in \cite{'tHV'79,Ball}). Following the notation
of refs.~\cite{JPA,UD1}, we write it as
\be
\label{defC1}
C^{(1)}(p_1^2, p_2^2, p_3^2) \equiv  J(4; 1, 1, 1)
= \mbox{i} \pi^2 \; (p_3^2)^{-1} \; \Phi^{(1)}(x,y),
\ee
\be
\label{xy}
 x \equiv {p_1^2}/{p_3^2} \hspace{0.5cm}\mbox{   ,   }
\hspace{0.5cm} y \equiv {p_2^2}/{p_3^2} \; .
\ee
The function $\Phi^{(1)}$ can be represented as a parametric integral,
\be
\label{Phi1int}
\vspace{-1mm}
\Phi^{(1)} (x,y) = - \Int_0^1
\frac{\mbox{d} \xi}{y \xi^2 + (1-x-y) \xi +x} \;
\left( \ln\frac{y}{x} + 2\ln{\xi}  \right),
\vspace{-3mm}
\ee
or in terms of dilogarithms,
\be
\label{Phi1}
\Phi^{(1)} (x,y) = \frac{1}{\lambda} \left\{ \frac{}{}
2 \left( \Li{2}{-\rho x} + \Li{2}{-\rho y} \right)
+ \ln\frac{y}{x} \ln{\frac{1+\rho y}{1+\rho x}}
+ \ln(\rho x) \ln(\rho y) + \frac{\pi^2}{3}
\right\} ,
\vspace{-1mm}
\ee
\be
\label{lambda}
\lambda(x,y) \equiv \sqrt{(1-x-y)^2 - 4 x y} \; \; \; ,
\; \; \; \rho(x,y) \equiv 2 \; (1-x-y+\lambda)^{-1} .
\ee

{\em The case} $\nu_1=\nu_2=\nu_3=1, \; n=4-2\ep$. In ref.~\cite{UD3},
the following representation was obtained for
arbitrary $n=4-2\ep$ :
\be
\label{int-dr}
\vspace{-1mm}
J(4-2\ep;1,1,1) = \frac{\pi^{2-\ep}\;\mbox{i}^{1+2\ep}}{(p_3^2)^{1+\ep}} \;
                 \frac{\Gamma (1+\ep) \Gamma^2 (1-\ep)}{\Gamma(1-2\ep)} \;
                 \frac{1}{\ep} \Int_0^1
 \frac{\mbox{d}\xi\;\xi^{-\ep} \left( (y\xi)^{-\ep} - (x/\xi)^{-\ep} \right)}
      {\left( y \xi^2 + (1-x-y)\xi + x \right)^{1-\ep}} .
\vspace{-2mm}
\ee
The expansion in $\ep$ yields
\be
\label{J111}
J(4-2\ep;1,1,1) = \pi^{2-\ep} \; \mbox{i}^{1+2\ep} (p_3^2)^{-1-\ep} \;
                 \Gamma (1+\ep) \;
    \left\{ \Phi^{(1)}(x,y) + \ep \; \Psi^{(1)}(x,y) + {\cal O}(\ep^2)
\right\},
\vspace{-1mm}
\ee
where $\Phi^{(1)}$ is defined by (\ref{Phi1int})--(\ref{Phi1})
while $\Psi^{(1)}$ can be represented as
\bea
\label{Psiint}
\vspace{-1mm}
\Psi^{(1)}(x,y) = - \Int_0^1 \frac{\mbox{d} \xi}
{y \xi^2 + (1 \! - \! x \! - \! y) \xi + x} \;
\left\{ \left( \ln\frac{y}{x} + 2\ln\xi \right)
        \ln(y \xi^2 + (1 \! - \! x \! - \! y) \xi + x) \right.
\nonumber \\[-2mm]
\left. -2 \ln y \ln\xi - 2 \ln^2\xi - \frac{1}{2} \ln(xy) \ln\frac{y}{x}
\right\} .  \hspace{1cm}
\vspace{-2mm}
\eea
The result in terms of polylogarithms (up to the
third order) is presented in \cite{UD3}, eq.~(29).

{\em The case} $\nu_1=\nu_2=1, \; \nu_3=1+\delta, \; n=4$.
In ref.~\cite{UD1},
the following formula was derived:
\be
\label{intPhi}
J(4; 1,1,1+\delta) = \frac{\mbox{i} \pi^2}{(p_3^2)^{1+\delta}} \;
\frac{1}{\delta} \Int_0^1 \mbox{d} \xi
\frac{(y \xi)^{-\delta} - (x/ \xi)^{-\delta}}
     {y \xi^2 + (1-x-y) \xi + x} .
\vspace{-1mm}
\ee
The first terms of the expansion of $J(4; 1,1,1+\delta)$ in $\delta$,
up to and including $\delta^2$, are
\bea
\label{expdelta}
\frac{\mbox{i} \pi^2}{(p_3^2)^{1+\delta}}
\left\{ \left( 1 \! - \frac{1}{2} \delta (\ln x \! + \! \ln y)
              + \frac{1}{6} \delta^2
                   (\ln^2 x \! + \! \ln x \ln y \! + \! \ln^2 y)
\right) \Phi^{(1)}(x,y)
+ \frac{1}{3} \delta^2  \Phi^{(2)}(x,y)
\right\}
\vspace{-2mm}
\eea
where $\Phi^{(2)}$ is connected with the two-loop
planar diagram (see in Section~3).

Note that in all integral representations (\ref{Phi1int}),
(\ref{int-dr}), (\ref{Psiint}) and (\ref{intPhi}) the denominator
of the integrand can be represented as a propagator,
$p_3^2 \; (y \xi^2 + (1-x-y) \xi + x) = (p_1 + \xi p_2)^2$.

\vspace{5mm}

{\bf 3} There are two basic ``topologies'' of the two-loop three-point
diagrams: planar (Fig.~1a) and non-planar (Fig.~1b) ones.
Note that the diagram in Fig.~1b is symmetric with respect to all
three external lines, while the diagram in Fig.~1a is symmetric
with respect to the two lower lines only.
The corresponding Feynman integrals are:
\be
\label{defC2}
C^{(2)}(p_1^2, p_2^2, p_3^2)
= \int\int \frac{\mbox{d} q \; \mbox{d} r}
                {(p_1+r)^2 \; (p_1+q)^2 \; (p_2-r)^2 \; (p_2-q)^2
                 \; r^2 \; (q-r)^2} ,
\ee
\be
\label{defCc2}
\widetilde{C}^{(2)}(p_1^2, p_2^2, p_3^2)
= \int\int \frac{\mbox{d} q \; \mbox{d} r}
                {(p_1+q)^2 \; (p_1+q+r)^2 \; (p_2-r)^2 \;
                   (p_2-q-r)^2 \;  r^2 \; q^2 }
\ee
\newcommand{\parta}
 {\setlength {\unitlength}{1.5mm}
 \begin{picture}(36,54)(0,0)
 \put (18,48) {\line(0,1){6}}
 \put (18,48) {\line(-1,-3){12}}
 \put (18,48) {\line(1,-3){12}}
 \put (13,33)  {\line(1,0){10}}
 \put (8,18)  {\line(1,0){20}}
 \put (18,48) {\circle*{1}}
 \put (13,33)  {\circle*{1}}
 \put (23,33) {\circle*{1}}
 \put (8,18)  {\circle*{1}}
 \put (28,18) {\circle*{1}}
 \put (1,15)  {\makebox(0,0)[bl]{\large $p_1$}}
 \put (32,15) {\makebox(0,0)[bl]{\large $p_2$}}
 \put (0,25)  {\makebox(0,0)[bl]{\large $p_1+r$}}
 \put (28,25) {\makebox(0,0)[bl]{\large $p_2-r$}}
 \put (6,43)  {\makebox(0,0)[bl]{\large $p_1+q$}}
 \put (22,43) {\makebox(0,0)[bl]{\large $p_2-q$}}
 \put (17,14)  {\makebox(0,0)[bl]{\large $r$}}
 \put (14,29) {\makebox(0,0)[bl]{\large $q-r$}}
 \put (20,51)  {\makebox(0,0)[bl]{\large $p_3$}}
 \put (15,3)  {\makebox(0,0)[bl]{\large (a)}}
 \end{picture}}
\newcommand{\partb}
 {\setlength {\unitlength}{1.5mm}
 \begin{picture}(26,54)(0,0)
 \put (18,48) {\line(0,1){6}}
 \put (18,48) {\line(-1,-3){12}}
 \put (18,48) {\line(1,-3){12}}
 \put (13,33)  {\line(1,-1){15}}
 \put (8,18)  {\line(1,1){15}}
 \put (18,48) {\circle*{1}}
 \put (13,33)  {\circle*{1}}
 \put (23,33) {\circle*{1}}
 \put (8,18)  {\circle*{1}}
 \put (28,18) {\circle*{1}}
 \put (1,15)  {\makebox(0,0)[bl]{\large $p_1$}}
 \put (32,15) {\makebox(0,0)[bl]{\large $p_2$}}
 \put (0,25)  {\makebox(0,0)[bl]{\large $p_1+q$}}
 \put (28,25) {\makebox(0,0)[bl]{\large $p_2-r$}}
 \put (1,43)  {\makebox(0,0)[bl]{\large $p_1+q+r$}}
 \put (22,43) {\makebox(0,0)[bl]{\large $p_2-q-r$}}
 \put (24,18)  {\makebox(0,0)[bl]{\large $r$}}
 \put (11,18) {\makebox(0,0)[bl]{\large $q$}}
 \put (20,51)  {\makebox(0,0)[bl]{\large $p_3$}}
 \put (15,3)  {\makebox(0,0)[bl]{\large (b)}}
 \end{picture}}
\begin{figure}[bth]
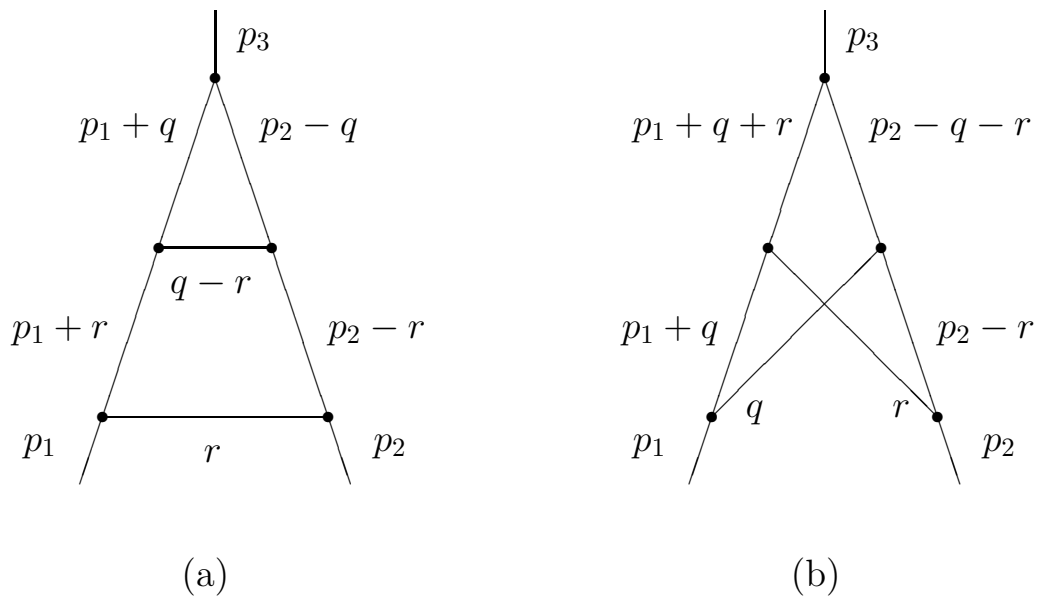

\[
\begin{array}{cccccc}
\parta & $\hspace{2cm}$ & \partb
\end{array}
\]
\caption{Two-loop three-point diagrams: planar (a) and
         non-planar (b) cases}
\end{figure}

The planar diagram (\ref{defC2}) was calculated (in four dimensions)
in ref.~\cite{UD1}. The result is
\be
\label{C2}
C^{(2)}(p_1^2, p_2^2, p_3^2) =
(\mbox{i} \pi^2)^2 \; (p_3^2)^{-2}
\; \Phi^{(2)}(x,y) ,
\ee
where the function $\Phi^{(2)}$ (see also (\ref{expdelta})) is
\be
\label{Phi2int}
\Phi^{(2)} (x,y) = - \frac{1}{2} \Int_0^1
\frac{\mbox{d} \xi}{y \xi^2 + (1-x-y) \xi +x} \;
\ln{\xi} \left( \ln\frac{y}{x} + \ln{\xi}  \right)
\left( \ln\frac{y}{x} + 2\ln{\xi}  \right) .
\ee
It can be expressed in terms of polylogarithms up to the fourth order,
see e.g. \cite{UD2}, eq.~(19)\footnote{We would like to note that
recently this result was checked numerically \cite{Czarn}.}.

As to the non-planar diagram (\ref{defCc2}),
in refs.~\cite{UD-YaF,UD3} it was obtained (also at $n=4$) that
\be
\label{Cc2}
\widetilde{C}^{(2)}(p_1^2, p_2^2, p_3^2)
= \left( C^{(1)}(p_1^2, p_2^2, p_3^2) \right) ^2
= (\mbox{i} \pi^2)^2 \; (p_3^2)^{-2} \;
  \left( \Phi^{(1)} (x,y) \right) ^2 ,
\ee
with $\Phi^{(1)} (x,y)$ defined by (\ref{Phi1int})--(\ref{Phi1}).
Therefore, $\widetilde{C}^{(2)}$ can be expressed
in terms of dilogarithms and their products.

Now, let us consider the cases when some numerators occur
in the integrands on r.h.s.'s of (\ref{defC2}), (\ref{defCc2}).
First of all, let us introduce the following notation:
\be
\label{C2num}
C^{(2)}\left[ \mbox{something} \right]
\equiv \left\{ \mbox{integral (\ref{defC2}) with}
        \; \left[ \mbox{something} \right] \;
       \mbox{in the numerator} \right\},
\ee
\be
\label{Cc2num}
\widetilde{C}^{(2)}\left[ \mbox{something} \right]
\equiv \left\{ \mbox{integral (\ref{defCc2}) with}
        \; \left[ \mbox{something} \right] \;
       \mbox{in the numerator} \right\} .
\ee
In this notation, eqs.~(\ref{defC2}) and (\ref{defCc2}) correspond
to $C^{(2)}[1]$ and $\widetilde{C}^{(2)}[1]$, respectively.
Inserting some numerators into
(\ref{defC2}) and (\ref{defCc2}) may produce divergent
integrals. In these cases, we shall understand that
dimensional regularization \cite{dimreg} is employed to regulate these
singularities\footnote{For simplicity, we put the dimensional
regularization scale parameter $\mu_0=1$.},
and (for two-loop integrals) we shall usually omit
terms vanishing as $\ep\to 0$.

In the paper \cite{UD3} some integrals of the type of (\ref{C2num}),
(\ref{Cc2num}) were considered, and the results can be
written (in the new notation) as
\be
\label{list1}
C^{(2)}\left[ (p_1 \! + \! r)^2 \right]
= \widetilde{C}^{(2)}\left[ (p_1 \! + \! q)^2 \right]
= \widetilde{C}^{(2)}\left[ q^2 \right]
= p_1^2 \; C^{(2)}(p_2^2,p_3^2,p_1^2)
= \frac{(\mbox{i} \pi^2)^2}{p_1^2} \;
  \Phi^{(2)}\left(\frac{1}{x}, \frac{y}{x}\right) ,
\ee
\be
\label{list1a}
\widetilde{C}^{(2)}\left[ (p_1+q+r)^2 \right]
= \widetilde{C}^{(2)}\left[ (p_2-q-r)^2 \right]
= p_3^2 \; C^{(2)}(p_1^2,p_2^2,p_3^2)
= \frac{(\mbox{i} \pi^2)^2}{p_3^2} \;
  \Phi^{(2)}(x,y) ,
\ee
\be
\label{list2}
C^{(2)}\left[ (q-r)^2 \right]
=\frac{\mbox{i}^{2+4\ep} \pi^{4-2\ep}}{(p_3^2)^{1+2\ep}} \;
 \frac{\Gamma^2(1+\ep)}{1-2\ep} \;
 \left\{ \frac{1}{\ep} \Phi^{(1)}(x,y) + \Psi^{(1)}(x,y) \right\},
\hspace{25mm}
\ee
\be
\label{list3}
C^{(2)}\left[ (p_1 \! + \! q)^2 \right]
=\frac{\mbox{i}^{2+4\ep} \pi^{4-2\ep}}{(p_1^2)^{1+2\ep}} \;
 \frac{\Gamma^2(1 \! +\! \ep)}{1-2\ep}
 \left\{ \left(\frac{1}{\ep} - \frac{1}{2} \ln\frac{y}{x^2} \right)
            \Phi^{(1)}\left(\frac{1}{x}, \frac{y}{x}\right)
          + \Psi^{(1)}\left(\frac{1}{x}, \frac{y}{x}\right) \right\},
\ee
\bea
\label{list4}
C^{(2)}\left[ (p_1+r)^2 (p_2-r)^2 \right]
= \widetilde{C}^{(2)}\left[ (p_1+q)^2 (p_2-r)^2 \right]
\hspace{55mm}
\nonumber \\
=\frac{\mbox{i}^{2+4\ep} \pi^{4-2\ep}}{(p_3^2)^{2\ep}} \;
 \frac{\Gamma^2(1+\ep)}{2 (1-2\ep) (1-3\ep)} \;
 \left\{ \frac{1}{\ep^2} - \ln x \ln y + (1-x-y) \Phi^{(1)}(x,y)
               - \frac{\pi^2}{3} \right\} .
\eea
One should remember that ``$+\cal{O}(\ep)$'' is understood
in all r.h.s.'s.
Some other results can be obtained from
(\ref{list1})--(\ref{list4}) by using symmetry properties,
or they may correspond to propagator-type diagrams,
for example (see \cite{zeta3})
\be
\label{6zeta3}
C^{(2)}\left[r^2\right] = (\mbox{i}\pi^2)^2 \; (p_3^2)^{-1} \; 6 \zeta(3) .
\ee

In all these examples (\ref{list1})--(\ref{6zeta3}) the numerators can
be cancelled against
the corresponding denominators, and (effectively) we obtain
diagrams where some of the lines are reduced to points.
There exist, however, some scalar numerators which cannot be
represented in terms of the denominators of the diagrams
in Fig.~1a,b; namely:
\be
\label{irred}
C^{(2)}\left[ q^2 \right] \hspace{1cm} \mbox{and}
\hspace{1cm} \widetilde{C}^{(2)}\left[ (q+r)^2 \right] .
\ee
Other irreducible cases can be related to these two ones.
It is easy to see that, if we introduce auxiliary
``forward scattering'' four-point functions according to
Fig.~2a and Fig.~2b, these numerators will correspond
to the dashed lines (missing in Fig.~1a and Fig.~1b).
Calculation of these integrals (\ref{irred})
will be considered in the next two sections.
\newcommand{\partaa}
 {\setlength {\unitlength}{1.5mm}
 \begin{picture}(46,54)(0,0)
 \put (13,10) {\line(0,1){44}}
 \put (33,10) {\line(0,1){44}}
 \multiput(13.5,48)(2,0){10}{\line(1,0){1}}
 \put (13,33) {\line(1,0){20}}
 \put (13,18) {\line(1,0){20}}
 \put (13,48) {\circle*{1}}
 \put (33,48) {\circle*{1}}
 \put (13,33) {\circle*{1}}
 \put (33,33) {\circle*{1}}
 \put (13,18) {\circle*{1}}
 \put (33,18) {\circle*{1}}
 \put (7,14)  {\makebox(0,0)[bl]{\large $p_1$}}
 \put (36,14) {\makebox(0,0)[bl]{\large $p_2$}}
 \put (7,51)  {\makebox(0,0)[bl]{\large $p_1$}}
 \put (36,51) {\makebox(0,0)[bl]{\large $p_2$}}
 \put (3,25)  {\makebox(0,0)[bl]{\large $p_1+r$}}
 \put (34,25) {\makebox(0,0)[bl]{\large $p_2-r$}}
 \put (3,43)  {\makebox(0,0)[bl]{\large $p_1+q$}}
 \put (34,43) {\makebox(0,0)[bl]{\large $p_2-q$}}
 \put (22,14) {\makebox(0,0)[bl]{\large $r$}}
 \put (19,29) {\makebox(0,0)[bl]{\large $q-r$}}
 \put (22,44) {\makebox(0,0)[bl]{\large $q$}}
 \put (20,3)  {\makebox(0,0)[bl]{\large (a)}}
 \end{picture}}
\newcommand{\partbb}
 {\setlength {\unitlength}{1.5mm}
 \begin{picture}(46,54)(0,0)
 \put (13,10) {\line(0,1){44}}
 \put (33,10) {\line(0,1){44}}
 \multiput(13.5,48)(2,0){10}{\line(1,0){1}}
 \put (13,33) {\line(4,-3){20}}
 \put (13,18) {\line(4,3){20}}
 \put (13,48) {\circle*{1}}
 \put (33,48) {\circle*{1}}
 \put (13,33) {\circle*{1}}
 \put (33,33) {\circle*{1}}
 \put (13,18) {\circle*{1}}
 \put (33,18) {\circle*{1}}
 \put (7,14)  {\makebox(0,0)[bl]{\large $p_1$}}
 \put (36,14) {\makebox(0,0)[bl]{\large $p_2$}}
 \put (7,51)  {\makebox(0,0)[bl]{\large $p_1$}}
 \put (36,51) {\makebox(0,0)[bl]{\large $p_2$}}
 \put (3,25)  {\makebox(0,0)[bl]{\large $p_1+q$}}
 \put (34,25) {\makebox(0,0)[bl]{\large $p_2-r$}}
 \put (-3,43)  {\makebox(0,0)[bl]{\large $p_1+q+r$}}
 \put (34,43) {\makebox(0,0)[bl]{\large $p_2-q-r$}}
 \put (17,17) {\makebox(0,0)[bl]{\large $r$}}
 \put (27,17) {\makebox(0,0)[bl]{\large $q$}}
 \put (19,44) {\makebox(0,0)[bl]{\large $q+r$}}
 \put (20,3)  {\makebox(0,0)[bl]{\large (b)}}
 \end{picture}}
\begin{figure}[bth]
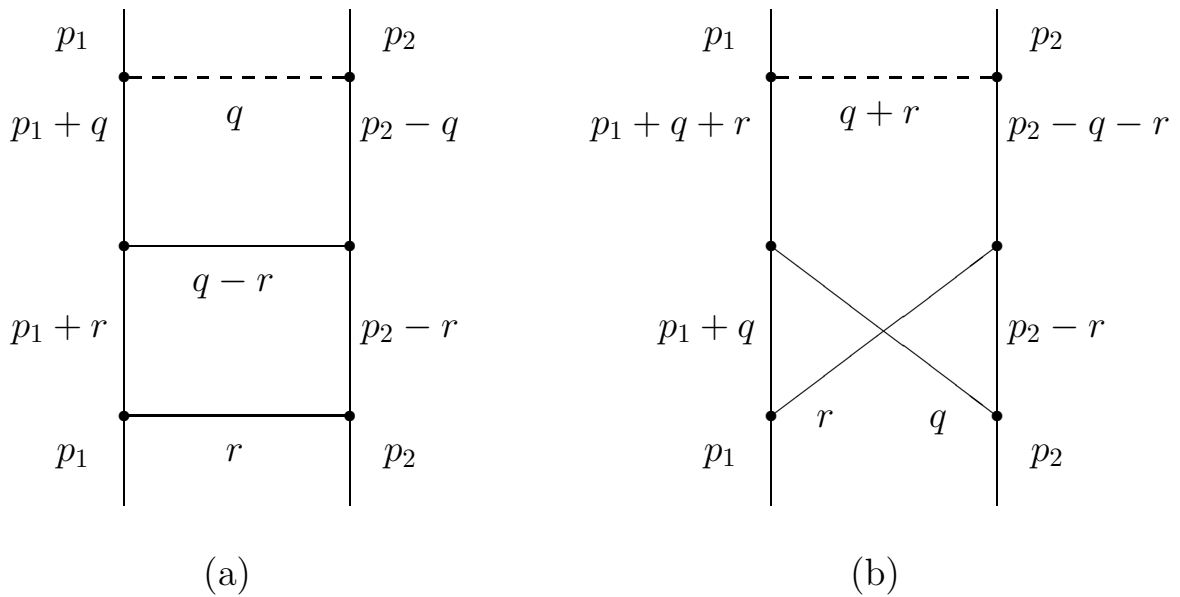

\[
\begin{array}{cccccc}
\partaa & $\hspace{1cm}$ & \partbb
\end{array}
\]
\caption{Auxiliary planar (a) and non-planar (b) four-point functions}
\end{figure}

\vspace{5mm}

{\bf 4.} In this section we shall examine the planar diagram
with irreducible numerator, $C^{(2)}\left[ q^2 \right]$.
It is easy to see that it can be written as
\be
\label{4-1}
C^{(2)}\left[ q^2 \right]
= C^{(2)}\left[ (p_1+q)^2 \right] - C^{(2)}\left[ (p_1+r)^2 \right]
  + C^{(2)}\left[ r^2 \right]
  - 2 p_1^{\mu} \; C^{(2)}\left[ (q-r)_{\mu} \right],
\ee
where only the first term on the r.h.s. is singular in $\ep$.
All the terms on the r.h.s. are known (see eqs.~(\ref{list4}),
(\ref{list1}), (\ref{6zeta3})), exept the last one,
$p_1^{\mu} \; C^{(2)}\left[ (q-r)_{\mu} \right]$, which is convergent.

Let us consider
$p_1^{\mu} \; C^{(2)}\left[ (q-r)_{\mu} \right]$ at $n=4$ and introduce
auxiliary analytic regularization by multiplying the integrand with
$ [(p_1+r)^2]^{-\delta} [(p_2-r)^2]^{-\delta} [r^2]^{\delta}
  [(q-r)^2]^{\delta}$. Since the integral is convergent,
the result should correspond
to the limit $\delta \to 0$. Then, let us consider $q$-integration
and integrate by parts\footnote{The calculation of
$C^{(2)}\left[ q^2 \right]$ can alternatively be
considered in the coordinate space.} (see in \cite{ibp}):
\bea
\label{ibp}
p_1^{\mu} \;
\int \frac{\mbox{d} q \; (q-r)_{\mu}}
          {(p_1+q)^2 \; (p_2-q)^2 \; [(q-r)^2]^{1-\delta}}
= \frac{1}{2\delta} \; p_1^{\mu} \;
\int \frac{\mbox{d} q }
          {(p_1+q)^2 \; (p_2-q)^2 }
\; \frac{\partial}{\partial q^{\mu}} [(q-r)^2]^{\delta}
\nonumber \\[1mm]
= \frac{1}{\delta}
\int \frac{\mbox{d} q \; (p_1, p_1+q)}
          {[(p_1+q)^2]^2 \; (p_2-q)^2 \; [(q-r)^2]^{-\delta}}
- \frac{1}{\delta}
\int \frac{\mbox{d} q \; (p_1, p_2-q)}
          {(p_1+q)^2 \; [(p_2-q)^2]^2 \; [(q-r)^2]^{-\delta}} .
\eea

Now, let us study the contributions corresponding to the two integrals
on the r.h.s. separately. Each of them is sungular in $\delta$, but
together they should give a finite result.

In the first contribution, the sum of the powers of
$(p_2-r)^2, (p_2-q)^2$ and $(q-r)^2$
in the denominator is equal to two (half the space-time dimension
if $n=4$). So, one can try to use the ``uniqueness'' relation
(\ref{uniq1}) to transform the product of these propagators
into a triangle. However, we need to introduce
an additional regularization by multiplying the
integrand with
$[(p_1+r)^2]^{-\delta'} [(p_2-q)^2]^{-\delta'} [(q-r)^2]^{\delta'}$
(otherwise we would obtain ``bare'' singularities in separate
terms). Since the contribution we are considering is
regular in $\delta'$ (but not in $\delta$), we need to remember
that, after the calculation is performed, we have to let
$\delta' \to 0$ first.
Now, we perform several transformations by using the
relations (\ref{uniq1}) and (\ref{uniq3}) (and also trivial formulae
for one-loop two-point functions), and we arrive at the
following representation of this contribution:
\bea
\label{firstc}
\frac{1}{\delta}
\int \int \frac{\mbox{d} q \; \mbox{d} r \; (p_1, p_1+q)}
          {[(p_1+r)^2]^{1+\delta+\delta'} \;
           [(p_1+q)^2]^2 \; [(p_2-r)^2]^{1+\delta} \;
           [(p_2-q)^2]^{1+\delta'} \; [r^2]^{1-\delta} \;
           [(q-r)^2]^{-\delta-\delta'}}
\nonumber \\
= \frac{\mbox{i} \pi^2}{2 \delta^2 (1+\delta)}
\left\{ - \mbox{i} \pi^2 \; (p_3^2)^{-1-\delta'} \; \frac{1}{\delta'}
        - (1+\delta') p_1^2 \; J(4; 1, 2+\delta', 1)
\right.
\hspace{4cm}
\nonumber \\
\left.
+\frac{1}{\delta'} (p_1^2)^{1+\delta}
\left[ (\delta + \delta')(1+ \delta + \delta') \;
                 J(4; 1, 2+\delta+\delta', 1)
        - \delta (1+ \delta) \; J(4; 1, 2+\delta, 1) \right]
\right\} ,
\eea
where the integrals $J$ are one-loop triangles defined by (\ref{defJ}).

Considering the second term on the r.h.s. of eq.~(\ref{ibp})
and using the fact that, due to momentum conservation,
$ -(p_1, p_2-q) = (p_2, p_2-q) + (p_3, p_2 - q)$,
it is easy to see that the term $(p_2, p_2-q)$ should give the
same as (\ref{firstc}) with $p_1 \leftrightarrow p_2$.
To calculate the term corresponding to $(p_3, p_2 - q)$,
we need to use only the relation (\ref{uniq1})
(and we do not need to introduce additional parameter $\delta'$).
In such a way, we get
\bea
\label{thirdc}
\frac{1}{\delta}
\int \int \frac{\mbox{d} q \; \mbox{d} r \; (p_3, p_2-q)}
          {[(p_1+r)^2]^{1+\delta} \;
           (p_1+q)^2 \; [(p_2-r)^2]^{1+\delta} \;
           [(p_2-q)^2]^2 \; [r^2]^{1-\delta} \;
           [(q-r)^2]^{-\delta}}
\nonumber \\
= - \frac{\mbox{i} \pi^2}{2 \delta (1+\delta)}
\; \left\{ J(4; 1,1,1) + (p_2^2)^{\delta} \; J(4; 1+ \delta, 1, 1) \right\}
\hspace{3cm}
\nonumber \\
- \frac{1}{2} \;
\int \int \frac{\mbox{d} q \; \mbox{d} r }
          {[(p_1+q)^2]^{1+\delta}  \; [(p_2-r)^2]^{1+\delta} \;
           [(p_2-q)^2]^{1-\delta} \; [r^2]^{1-\delta} \;
           (q-r)^2} .
\eea
Note that in the last integral on the r.h.s. the denominator
$(p_1+r)^2$ is missing. As $\delta \to 0$, this integral
gives $C^{(2)}\left[ (p_1+r)^2 \right]$ which cancels the
corresponding term in (\ref{4-1}).

Careful analysis of the contributions (\ref{firstc}), (\ref{thirdc})
shows that what we need, in addition to the formulae presented
in section~2, is the expansion of the integral $J(4; 1, 1, 2+\delta)$
up to (and including) $\delta^2$ terms. To get it, it is
convenient to use the representation (\ref{intPhi}) with
$\delta$ substituted by $1+\delta$. The result of this calculation
can be presented as
\bea
\label{aOmega}
J(4; 1, 1, 2+\delta) =
\frac{\mbox{i} \pi^2}{(p_3^2)^{2+\delta}}
 \frac{1}{x y} \; \frac{1}{1+\delta}
\left\{
- \frac{1}{\delta} + (\ln x + \ln y)
- \frac{\delta}{2} \left(\ln^2 x + \ln x \ln y + \ln^2 y \right)
\right.
\nonumber \\
+ \frac{\delta^2}{6}
     \left(\ln^3 x + \ln^2 x \ln y + \ln x \ln^2 y + \ln^3 y \right)
\hspace{20mm}
\nonumber \\
\left.
- \frac{\delta}{2} \;
\left( 1 - \frac{\delta}{2} (\ln x + \ln y) \right)
(1-x-y) \; \Phi^{(1)}(x,y)
+ \frac{\delta^2}{6} \; \Omega^{(2)}(x,y) + {\cal O}(\delta^3)
\right\} .
\eea
Here we define a set of functions $\Omega^{(N)}$
via the derivatives of the functions $\Phi^{(N)}$
(the general case of these functions was considered in \cite{UD2}) as
\be
\label{OmegaPhiN}
\Omega^{(N)}(x,y) = \lambda \;
\left[ x \frac{\partial}{\partial x} \left(\lambda \Phi^{(N)}(x,y) \right)
     + y \frac{\partial}{\partial y} \left(\lambda \Phi^{(N)}(x,y) \right)
\right] .
\ee

For $N=1$, the corresponding function $\Omega^{(1)}$ is trivial,
\be
\label{Omega1}
\Omega^{(1)}(x,y) = \ln x + \ln y - (x-y) \ln\frac{y}{x} .
\ee
For $N=2$, the function $\Phi^{(2)}$ is defined by eq.~(\ref{Phi2int}),
and $\Omega^{(2)}$ can be represented as
\be
\label{Omega2Int}
\Omega^{(2)}(x,y) = \frac{1}{2}\ln x \ln^2\frac{y}{x}
  + 3 \Int_0^1 \frac{\mbox{d} \xi}{\xi}
    \left(\ln\frac{y}{x} + 2 \ln\xi\right)
    \ln\left(\frac{y\xi^2 + (1-x-y)\xi + x}{x}\right),
\ee
or in terms of polylogarithms,
\bea
\label{Omega2}
\Omega^{(2)}(x,y) = 6 \left[ \Li{3}{-\rho x} + \Li{3}{-\rho y} \right]
              + 3 \ln\frac{y}{x}
                \left[ \Li{2}{-\rho x} - \Li{2}{-\rho y} \right]
\hspace{18mm}
\nonumber \\
              - \frac{1}{2} \ln^2\frac{y}{x}
                \left[\ln(1+\rho x) + \ln(1+\rho y) \right]
              + \frac{1}{2} \left( \pi^2 + \ln(\rho x) \ln(\rho y) \right)
                \left[ \ln(\rho x) + \ln(\rho y) \right] .
\eea
Note that, combining the derivatives of $\Phi^{(2)}$ in a different
way, we may obtain $\Phi^{(1)}$ as
\be
\label{Phi1Phi2}
\ln\frac{y}{x} \; \Phi^{(1)}(x,y) = \frac{2}{\lambda} \;
\left[ (1 \! - \! x \! + \! y) x \frac{\partial}{\partial x}
               \left(\lambda \Phi^{(2)}(x,y) \right)
     - (1 \! + \! x \! - \! y) y \frac{\partial}{\partial y}
               \left(\lambda \Phi^{(2)}(x,y) \right)
\right] .
\ee

Finally, using (\ref{aOmega}), the integral with irreducible
numerator can be represented as
\bea
\label{planar}
C^{(2)}\left[ q^2 \right]
= C^{(2)}\left[ (q-r)^2 \right] + C^{(2)}\left[ r^2 \right]
- \frac{(i\pi^2)^2}{4 \; p_3^2} \;
\left[ \ln x + \ln y - (x-y)\ln\frac{y}{x} \right] \; \Phi^{(1)}(x,y)
\nonumber \\
- \frac{(i\pi^2)^2}{2 \; p_3^2} \;
\left[ \Omega^{(2)}\left( \frac{x}{y}, \frac{1}{y} \right)
      + \Omega^{(2)}\left( \frac{y}{x}, \frac{1}{x} \right) \right]
+ \cal{O}(\ep) .
\eea
Note that the factor multiplying $\Phi^{(1)}$ is proportional to
$\Omega^{(1)}(x,y)$, eq.~(\ref{Omega1}).

\vspace{5mm}

{\bf 5.} Let us consider the non-planar diagram
with irreducible numerator,
$\widetilde{C}^{(2)}\left[ (q+r)^2 \right]$. It is easy to see that
it is convergent and can be written as
\be
\widetilde{C}^{(2)}\left[ (q+r)^2 \right]
= p_2^2 \widetilde{C}^{(2)}\left[ 1 \right]
+ \widetilde{C}^{(2)}\left[ (p_2-q-r)^2 \right]
- 2 \widetilde{C}^{(2)}\left[ (p_2, p_2-q-r) \right] .
\ee

Now, using the symmetry of the non-planar diagram it is possible
to see that\footnote{This can be shown using
the fact that the non-planar integral
(\ref{defCc2}) remains invariant when we simultaneously change
$r \to p_2 - r$ and $q \to -p_1-q$.}
\be
\label{sym}
\widetilde{C}^{(2)}\left[ (p_2, p_2-q-r) \right]
= \widetilde{C}^{(2)}\left[ (p_2, p_1+q+r) \right] .
\ee
Writing the l.h.s. of (\ref{sym}) as half the sum
of the l.h.s. and the r.h.s., we get
\be
\widetilde{C}^{(2)}\left[ (p_2, p_2-q-r) \right]
= \frac{1}{2} \; (p_2, p_1+p_2) \; \widetilde{C}^{(2)}\left[ 1 \right] ,
\ee
and the result for the considered integral reduces to
\be
\widetilde{C}^{(2)}\left[ (q+r)^2 \right]
= \widetilde{C}^{(2)}\left[ (p_2-q-r)^2 \right]
- (p_1 p_2) \widetilde{C}^{(2)}\left[ 1 \right] .
\ee

According to (\ref{list1}) (see also \cite{UD3}),
$\widetilde{C}^{(2)}\left[ (p_2-q-r)^2 \right]$
can be represented in terms of the planar diagram as
$p_3^2 \; C^{(2)}\left[ 1 \right]$.
So, finally we arrive at the following result:
\be
\label{result2}
\widetilde{C}^{(2)}\left[ (q+r)^2 \right]
= p_3^2 \; C^{(2)}(p_1^2, p_2^2, p_3^2)
- (p_1 p_2) \; \widetilde{C}^{(2)}(p_1^2, p_2^2, p_3^2)
\ee
where the integrals contributing to the r.h.s. are defined
by eqs.~(\ref{defC2}) and (\ref{defCc2}).
This equation is illustrated by Fig.~3 where ``$-1$'' means that
the irreducible numerator can be considered as a negative
power of the corresponding denominator.
It is interesting that we have a ``mixture''
of different topologies on the r.h.s. of
eq.~(\ref{result2}) and Fig.~3.
\newcommand{\partaaa}
 {\setlength {\unitlength}{1mm}
 \begin{picture}(36,54)(0,0)
 \put (18,48) {\line(0,1){6}}
 \put (18,48) {\line(-1,-3){12}}
 \put (18,48) {\line(1,-3){12}}
 \put (13,33)  {\line(1,0){10}}
 \put (8,18)  {\line(1,0){20}}
 \put (18,48) {\circle*{1}}
 \put (13,33)  {\circle*{1}}
 \put (23,33) {\circle*{1}}
 \put (8,18)  {\circle*{1}}
 \put (28,18) {\circle*{1}}
 \put (1,15)  {\makebox(0,0)[bl]{\large $p_1$}}
 \put (32,15) {\makebox(0,0)[bl]{\large $p_2$}}
 \put (20,51)  {\makebox(0,0)[bl]{\large $p_3$}}
 \end{picture}}
\newcommand{\partbbb}
 {\setlength {\unitlength}{1mm}
 \begin{picture}(40,54)(0,0)
 \put (7,10) {\line(0,1){44}}
 \put (27,10) {\line(0,1){44}}
 \multiput(7.5,48)(2,0){10}{\line(1,0){1}}
 \put (7,33) {\line(4,-3){20}}
 \put (7,18) {\line(4,3){20}}
 \put (7,48) {\circle*{1}}
 \put (27,48) {\circle*{1}}
 \put (7,33) {\circle*{1}}
 \put (27,33) {\circle*{1}}
 \put (7,18) {\circle*{1}}
 \put (27,18) {\circle*{1}}
 \put (1,14)  {\makebox(0,0)[bl]{\large $p_1$}}
 \put (30,14) {\makebox(0,0)[bl]{\large $p_2$}}
 \put (1,51)  {\makebox(0,0)[bl]{\large $p_1$}}
 \put (30,51) {\makebox(0,0)[bl]{\large $p_2$}}
 \put (14,43) {\makebox(0,0)[bl]{$-1$}}
 \end{picture}}
\newcommand{\partccc}
 {\setlength {\unitlength}{1mm}
 \begin{picture}(26,54)(0,0)
 \put (18,48) {\line(0,1){6}}
 \put (18,48) {\line(-1,-3){12}}
 \put (18,48) {\line(1,-3){12}}
 \put (13,33)  {\line(1,-1){15}}
 \put (8,18)  {\line(1,1){15}}
 \put (18,48) {\circle*{1}}
 \put (13,33)  {\circle*{1}}
 \put (23,33) {\circle*{1}}
 \put (8,18)  {\circle*{1}}
 \put (28,18) {\circle*{1}}
 \put (1,15)  {\makebox(0,0)[bl]{\large $p_1$}}
 \put (32,15) {\makebox(0,0)[bl]{\large $p_2$}}
 \put (20,51)  {\makebox(0,0)[bl]{\large $p_3$}}
 \end{picture}}
\newcommand{\extraa}
{\setlength {\unitlength}{1mm}
\begin{picture}(15,54)(0,0)
 \put (0,30) {\makebox(0,0)[bl]{\Large$= \hspace{5mm} p_3^2$}}
 \end{picture}}
\newcommand{\extrab}
{\setlength {\unitlength}{1mm}
\begin{picture}(25,54)(0,0)
 \put (0,30) {\makebox(0,0)[bl]{\Large$ - \hspace{5mm} (p_1 p_2)$}}
 \end{picture}}
\begin{figure}[bth]
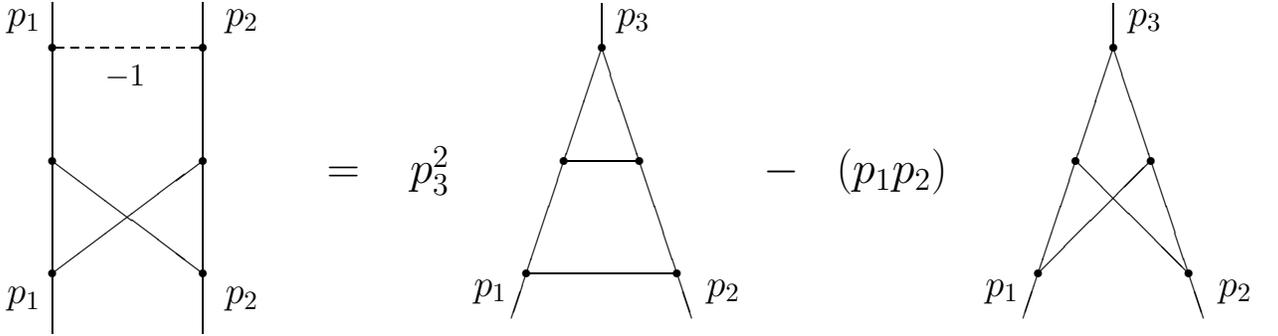

\[
\begin{array}{ccccc}
\partbbb & \extraa & \partaaa &
\extrab &
\partccc
\end{array}
\]
\caption{Result for the non-planar diagram with irreducible numerator}
\end{figure}

\vspace{5mm}

{\bf 6.} In this section we shall demonstrate that the problem
of tensor decomposition of two-loop three-point functions also
requires integrals with irreducible numerators.

Let us consider the case with one free Lorentz index.
It is easy to show \cite{BF,PV} that
\be
\label{vector}
C^{(2)}\left[ q_{\mu} \right] = \Delta^{-1}
\left\{ \left( p_2^2 {p_1}_{\mu} - (p_1 p_2) {p_2}_{\mu} \right)
         C^{(2)}\left[ (p_1 q) \right]
       +\left( p_1^2 {p_2}_{\mu} - (p_1 p_2) {p_1}_{\mu} \right)
         C^{(2)}\left[ (p_2 q) \right] \right\}
\ee
where $\Delta$ is nothing but the K{\"a}llen function,
$\Delta = p_1^2 p_2^2 - (p_1 p_2)^2 = (p_3^2)^2 \lambda^2(x,y)$ .
The same as (\ref{vector}) is valid for
$\widetilde{C}^{(2)}\left[ q_{\mu} \right]$
(we only need to change $C^{(2)} \to \widetilde{C}^{(2)}$
on the r.h.s.). Moreover, the results for $C^{(2)}\left[ r_{\mu} \right]$
and $\widetilde{C}^{(2)}\left[ r_{\mu} \right]$ can be obtained
by changing $q$  into $r$ everywhere.

For the two-loop planar diagram, we express $(p_1 q)$ and $(p_2 q)$
in terms of the denominators of the diagram in Fig.~2a,
and we get irreducible numerators (\ref{irred}),
\be
\label{C2p1q}
C^{(2)}\left[ (p_1 q) \right] = \frac{1}{2}
\left\{ - p_1^2 C^{(2)}\left[ 1 \right]
        + C^{(2)}\left[ (p_1+q)^2 \right]
        - C^{(2)}\left[ q^2 \right]
\right\} ,
\ee
\be
\label{C2p2q}
C^{(2)}\left[ (p_2 q) \right] = \frac{1}{2}
\left\{ p_2^2 C^{(2)}\left[ 1 \right]
        - C^{(2)}\left[ (p_2-q)^2 \right]
        + C^{(2)}\left[ q^2 \right]
\right\} .
\ee
Note that in $C^{(2)}\left[ (p_3 q) \right]$ the irreducible
integrals $C^{(2)}\left[ q^2 \right]$ disappear.
In the case of $C^{(2)}\left[ r_{\mu} \right]$, we
do not have irreducible numerators either, because $q^2$ in
(\ref{C2p1q})--(\ref{C2p2q})
should be replaced by $r^2$, and the corresponding integral
(\ref{6zeta3}) is trivial.

For the non-planar case, we need to express $(p_1 q)$ and $(p_2 q)$
in terms of denominators of the crossed diagram shown in Fig.~2b.
So, we get
\be
\label{Cc2p1q}
\widetilde{C}^{(2)}\left[ (p_1 q) \right] = \frac{1}{2}
\left\{- p_1^2 \widetilde{C}^{(2)}\left[ 1 \right]
       + \widetilde{C}^{(2)}\left[ (p_1+q)^2 \right]
       - \widetilde{C}^{(2)}\left[ q^2 \right]
\right\} ,
\ee
\be
\label{Cc2p2q}
\widetilde{C}^{(2)}\left[ (p_2 q) \right] = \frac{1}{2}
\left\{- \widetilde{C}^{(2)}\left[ (p_2 \! - \! q \! - \! r)^2 \right]
       + \! \widetilde{C}^{(2)}\left[ (p_2 \! - \! r)^2 \right]
       - \! \widetilde{C}^{(2)}\left[ r^2 \right]
       + \! \widetilde{C}^{(2)}\left[ (q \! + \! r)^2 \right]
\right\} ,
\ee
and we see that the irreducible numerator $(q+r)^2$ appears in
(\ref{Cc2p2q}). Analogous results for
$\widetilde{C}^{(2)}\left[ (p_1 r) \right]$
and $\widetilde{C}^{(2)}\left[ (p_2 r) \right]$
can be obtained from (\ref{Cc2p1q})--(\ref{Cc2p2q}) by using the
symmetry of the non-planar diagram.

When we consider more free Lorentz indices,
$C^{(2)}\left[ q_{\mu} q_{\sigma} \right],
C^{(2)}\left[ q_{\mu} r_{\sigma} \right]$, etc., we need
integrals with higher powers of irreducible numerators.
Such integrals will be considered in more detail in \cite{DOT}.

\vspace{5mm}

{\bf 7.} In the present paper we have considered an approach to the
calculation of three-point two-loop diagrams (Fig.~1a,b) with irreducible
numerators (\ref{irred}) which correspond to the dashed lines of
auxiliary four-point diagrams shown in Fig.~2a,b. We have obtained
exact results for both planar (Fig.~1a) and non-planar (Fig.~1b)
cases. The result for the non-planar case can be expressed in terms
of the functions corresponding to the diagrams without numerators,
as shown in Fig.~3. At the same time, the planar case is more
complicated, and the result (\ref{planar}) contains a new function
$\Omega^{(2)}$ which can be represented in terms of the derivatives
of the function $\Phi^{(2)}$ (see (\ref{OmegaPhiN})). So, the full set of
``non-trivial'' functions occurring in two-loop three-point
calculations (with massless internal particles and off-shell
external momenta) are: \\
$\Phi^{(1)}$, eqs.~(\ref{Phi1int})--(\ref{Phi1}), corresponding to the
one-loop triangle in four dimensions (\ref{defC1}); \\
$\Psi^{(1)}$, eq.~(\ref{Psiint}) (and eq.~(29) of \cite{UD3}), $\ep$-part of
the one-loop triangle;\\
$\Phi^{(2)}$, eqs.~(\ref{Phi2int}) (and eq.~(19) of \cite{UD2}),
corresponding to the two-loop planar diagram; \\
$\Omega^{(2)}$, eqs.~(\ref{Omega2Int})--(\ref{Omega2}), arising in the
two-loop planar diagram with irreducible numerator. \\
It is interesting that all these functions can be obtained from
expansions of dimensionally or analytically regularized
one-loop integrals in the regulating parameters ($\ep$ or $\delta$),
see eqs. (\ref{J111}), (\ref{expdelta}), (\ref{aOmega}).
All of them can be expressed in terms of polylogarithms.
Note that the results for some other diagrams can also be expressed
in terms of these functions, for example: one- and two-loop off-shell
four-point functions with massless internal lines \cite{UD1},
one-loop diagrams with higher number of external lines \cite{pent},
two-loop massive vacuum diagrams \cite{BDST,dual}.

We have also discussed the problem of tensor reduction of two-loop
three-point integrals which also requires integrals with irreducible
numerators. A recursive procedure for calculating integrals with
higher powers of irreducible numerators can be constructed \cite{DOT}.
Such integrals are needed for the decomposition of tensor integrals with
a larger number of uncontracted Lorentz indices.

Important applications of the obtained results are connected with
the large momentum expansion
of two-loop three-point massive diagrams, and with the calculation
of vertex corrections in massless QCD.

\vspace{2mm}

{\bf Acknowledgements.}
We are grateful to P.~Osland for his interest in this work,
useful discussions and help. In fact, during the work of one
of the authors, A.~D., with P.~Osland and O.V.~Tarasov \cite{DOT}, some
of the results presented in this paper were re-calculated
and cross-checked.
A.~D. is thankful to F.A.~Berends for discussions on the importance of
this activity for asymtotic expansions of three-point
two-loop diagrams with masses. We are also grateful to D.J.~Broadhurst
for useful comments about the problem of irreducible numerators.
The research was supported by the Research Council of Norway.

\vspace{5mm}

\end{document}